\documentclass[preprint2]{aastex}
\usepackage{lscape}
\usepackage{color}
\usepackage{subfigure}
\def\mc{$\mu$m}

\slugcomment{Not to appear in Nonlearned J., 45.}

\shorttitle{Mid-IR spectroscopy NGC\,3281}
\shortauthors{Sales et al.}

\begin{document}


\title{THE COMPTON-THICK SEYFERT 2 NUCLEUS OF NGC\,3281: TORUS CONSTRAINTS FROM THE 9.7\mc\ SILICATE ABSORPTION}


\author{Dinalva A. Sales\altaffilmark{1}; M. G. Pastoriza\altaffilmark{1,2}; R. Riffel\altaffilmark{1}; C. Winge\altaffilmark{3}; A. Rodr{\'{\i}}guez-Ardila\altaffilmark{4}; A. C. Carciofi\altaffilmark{5}}
\altaffiltext{1}{Departamento de Astronomia, Universidade Federal do Rio Grande do Sul. Av. Bento Gon\c calves 9500, Porto Alegre, RS, Brazil}
\altaffiltext{2}{Conselho Nacional de Desenvolvimento Cient\' ifico e Tecnol\' ogico, Brazil}
\altaffiltext{3}{Gemini Observatory, c/o Aura, Inc., Casilla 603, La Serena, Chile}
\altaffiltext{4}{Laborat\' orio Nacional de Astrof\' isica/MCT, Rua dos Estados Unidos 154, Itajub\' a, MG, Brazil}
\altaffiltext{5}{Instituto de Astronomia, Geof\' isica e Ci\^{e}ncias Atmosf\' ericas, Universidade de S\~ ao Paulo, Rua do Mat\~ ao 1226, Cidade Universit\' aria, S\~ ao Paulo, SP, Brazil}
\email{dinalva.aires@ufrgs.br}



\begin{abstract}

We present mid infrared (Mid-IR) spectra of the
Compton-thick Seyfert 2 galaxy NGC\,3281, obtained with the 
Thermal-Region Camera Spectrograph (T-ReCS) at the
Gemini South telescope. The spectra present a very deep silicate absorption at 9.7\,$\mu$m, and [S{\sc\,iv]}\,10.5\,$\mu$m and [Ne{\sc\,ii]}\,12.7\,$\mu$m ionic lines,
but no evidence of PAH emission. 
We find that the nuclear optical extinction is in the range 24 $\leq$ A$_{V}$ $\leq$ 83\,mag.
A temperature T = 300\,K was found for the black-body dust continuum component
of the unresolved 65\,pc nucleus and at 130\,pc SE, while the region at
130\,pc reveals a colder temperature (200\,K). 
We describe the nuclear spectrum of NGC\,3281 using a clumpy torus model 
that suggests that the nucleus of this galaxy hosts a dusty toroidal structure. 
According to this model, the ratio between the inner and outer radius
of the torus in NGC\,3281 is $R_0/R_d$ = 20, 
with 14 clouds in the equatorial radius 
with optical depth of
$\tau_{V}$ = 40\,mag. We would be looking in the direction of the 
torus equatorial radius ($i$ = 60$^{\circ}$), which has outer radius of R$_{0}\,\sim$ 11\,pc.
The column density is N$_{H}\approx$\,1.2\,$\times\,10^{24}\,cm^{-2}$ and iron K$\alpha$ equivalent width ($\approx$  0.5 - 1.2\,keV) are used to
check the torus geometry.
Our findings indicate that the X-ray absorbing column density, which classifies NGC\,3281 as a Compton-thick source, may also be responsible for the absorption at 9.7\,$\mu$m
providing strong evidence that the silicate dust responsible for this absorption can be located in the AGN torus.

\end{abstract}





\section{Introduction}

Mid-infrared (Mid-IR) spectra of active galactic nuclei (AGNs) are rich in emission features, such as polycyclic aromatic hydrocarbons (PAHs), molecular hydrogen and prominent forbidden emission lines \citep[]{sturm00,weedman05,wu09,gallimore10,sales10}. Another spectral feature commonly observed in AGNs is silicate at $\sim$ 9.7\,$\mu$m and 18\,$\mu$m, both in emission and absorption.

The AGN unified model proposes the existence of a dense concentration of absorbing material surrounding the central engine  in a toroidal distribution, which blocks the broad line region (BLR) emission from the line of sight in Seyfert 2 (Sy\,2) galaxies \citep{antonucci93}. However, it is not yet clear if the molecular gas and dust detected in Mid-IR spectra is actually associated with the absorbing material required by the so-called torus in the unified model.  

\citet{hao07} studied the distribution of the silicate feature strengths in a sample of 196 AGNs and ultraluminous infrared galaxies (ULIRGs). They found that the spectra of quasars and Seyfert~1 (Sy\,1) galaxies are characterized by silicate features in emission, with few Sy\,1s presenting weak absorptions. In contrast, Sy\,2 spectra are dominated by weak silicate absorption. These results suggest that the silicate emission (or absorption) is related to the line of sight, in the framework of the AGN unified model. In addition, most of the ULIRGs show very deep silicate absorption \citep{hao07}.

The silicate behavior in AGNs can be explained using Nenkova's torus models \citep{nenkova02,nenkova08a,nenkova08b}, which consider an ensemble of individual clouds instead of the standard ``solid doughnut''. \citet{mason06} detected the silicate emission feature in the Sy\,2 galaxy NGC\,2110, and compared it with  the \citet{nenkova08b} clumpy torus models, demonstrating that the presence of silicate emission can also be explained by an edge-on clumpy distribution. From their best-fit model, they conclude that there would be a small number of clouds along the line of sight, with a distribution that does not extend far above the torus equatorial plane.

According to \citet{hao07} the presence of strong silicate absorption and the absence of PAH features in some active galaxies, can be interpreted as extremely heavily obscuration in these sources. An example of such an object is NGC\,3281. Its spectrum shows a very deep silicate absorption at 9.7\,$\mu$m, as well as the forbidden emission lines [S{\sc\,iii]}\,18.7\,$\mu$m, [S{\sc\,iv]}\,10.5\,$\mu$m, [Ne{\sc\,ii]}\,12.8\,$\mu$m, [Ne{\sc\,iii]}\,15.5\,$\mu$m, [Ne{\sc\,v]}\,14.3\,$\mu$m and 24.3\,$\mu$m, [Ne{\sc\,vi]}\,7.6\,$\mu$m and [O{\sc\,iv]}\,25.8\,$\mu$m. No PAH emission was detected in the Spitzer spectrum \citep{wu09}. In addition, NGC\,3281 is a luminous infrared galaxy \citep[log\,L$_{IR} \sim$ 10.73\,L$_\odot$,][]{sanders03}, classified as a Sy\,2. Its optical spectrum was studied by \citet{thaisa92}, finding that the kinematics of the kpc extended ionized gas can be described by rotation in a plane, with material outflowing ($v\sim$\,150 km s$^{-1}$) from the nucleus within a conical structure. They also suggest that NGC\,3281 has dust asymmetrically distributed with respect to the ionized gas. 

In addition, using Advanced Satellite for Cosmology and Astrophysics (\textit{ASCA}) data,  \citet{simpson98} found that NGC\,3281 shows an extremely complex X-ray spectrum, with a prominent iron emission line and a large absorbing column density (N$_{H} \sim$ 7.1\,$\pm$\,1.2 $\times$ 10$^{23}$\,cm$^{-2}$) that when combined with the nuclear extinction inferred from the near-infrared color analysis (A$_{\textit{\sc v}}$ = 22\,$\pm$\,11\,mag), results in a N$_{H}$/A$_{\textit{\sc v}}$ ratio one order of magnitude higher than that of the Galaxy \citep{bohlin78}. This effect was attributed to an optically thick material along the line of sight of both the X-ray and the infrared emitting regions, which obscures the entire X-ray source but only a fraction of the much larger IR emitting region. Later, \citet{vignali02} using the Italian-Dutch satellite \textit{BeppoSAX}, found a N$_{H}$/A$_{\textit{\sc v}}$ ratio about twice that derived by \citet{simpson98}, adding NGC\,3281 to the list of AGNs presenting dust properties different from the Galaxy. The difference in the N$_{H}$/A$_{\textit{\sc v}}$ values derived by both studies, according to \citet{vignali02}, is due to the limited \textit{ASCA}  band-pass. \citet{vignali02} also show that its X-ray spectrum (E\,$<$\,10\,keV) is reflection-dominated, with a relatively strong K$\alpha$ iron emission line and a heavily absorbed nuclear continuum (N$_{H}\,\simeq\,1.5-2\,\times$ 10$^{24}$\,cm$^{-2}$). Hence, they classify NGC\,3281 as a Compton-thick galaxy. The high values found for the absorption column density in NGC\,3281 also explain the extremely high ratio F$_{[O{\sc\,III]\,}}$/F$_{2-10\,keV}$ $\approx$ 0.3, when compared to the average ($\approx$ 0.02) value obtained for a sample of Sy\,2 galaxies \citep{mulchaey94}.

It follows from the above discussion that this galaxy has a very heavily obscured nucleus, which makes it a key object to study whether or not the observed dust absortion is 
associated with the dusty torus of the unified model.
Thus, it may provide unambiguous observational evidence of this structure.
We present in this paper ground bases Mid-IR high-angular resolution spectra of the galaxy NGC\,3281. The Thermal-Region Camera Spectrograph \citep[T-ReCS;][]{telesco98} attached to the Gemini South 
telescope give a spatial resolution of 16 pc/pix, which is very adequate to study the dust distribution in the $\sim$ 200\,pc central radius of this galaxy. 
This paper is structured as follows: in Section \ref{observation} we describe the observation and data reduction processes.
Results are discussed in Section \ref{results}. Summary and conclusion are presented in Section \ref{conclusions}.

\section{Observations and Data Reduction}\label{observation}

Ground-based Mid-IR observations of NGC\,3281 were obtained with the T-ReCS in queue mode at Gemini South, in 2009 April 06 UT, as part of program GS-2009A-Q-34. Conditions were clear, with water vapor column in the range 5--8mm, and image quality of 0\farcs33 in the N band, as measured from the acquisition images of the telluric standards observed right before and/or after the science frames. All observations used a standard chop-nod technique to remove time-variable sky background, telescope thermal emission, and the effect of 1/\textit{f} noise from the array/electronics. The slit was oriented along PA =  315\arcdeg, with a chop throw of 15\arcsec, oriented along PA = 225\arcdeg, perpendicular to the slit direction, in order to include only the signal of the guided beam position in the frame and avoid possible nod effects in the spatial direction. The same slit position/nod orientation were used for the telluric standards. The instrument configuration used the low-resolution (R $\sim$ 100) grating and the 0\farcs36 slit,  for a dispersion of 0.0223\,$\mu$m/pixel and a spectral resolution of 0.08\,$\mu$m. The pixel scale is 0.089 arcsec/pixel in the spatial direction, and the slit is 21\farcs6 long. Spectral coverage was of 7.1$\mu$m centered at 10.5$\mu$m. The total on-source integration time was 980 sec.

The data reduction was performed using the {\sc midir} and {\sc gnirs} sub-packages of Gemini {\sc iraf}\footnote{IRAF is distributed by the National Optical Astronomy Observatory, which is operated by the Association of Universities for Research in Astronomy (AURA) under cooperative agreement with the National Science Foundation.} package. To extract the final spectrum,  we combined the chop- and nod-subtracted frames using the tasks {\sc tprepare} and {\sc mistack} in the {\sc midir} package. Wavelength calibration was obtained from the sky lines in the raw frames. To remove the telluric absorption lines from the galaxy spectrum, we divided the extracted spectrum for each observing night by that of the telluric standard stars HD\,3438 or HD\,4786 \citep{cohen99}, observed before and/or after the science target. Finally, the spectrum was flux calibrated by interpolating a black-body function to the spectrum of the telluric standard. For this step we used the task {\sc mstelluric} in the {\sc midir} package.

\begin{figure*}
 \begin{centering}
  \begin{tabular}{cc}
  \subfigure[ ]{\includegraphics[clip=true,width=8cm]{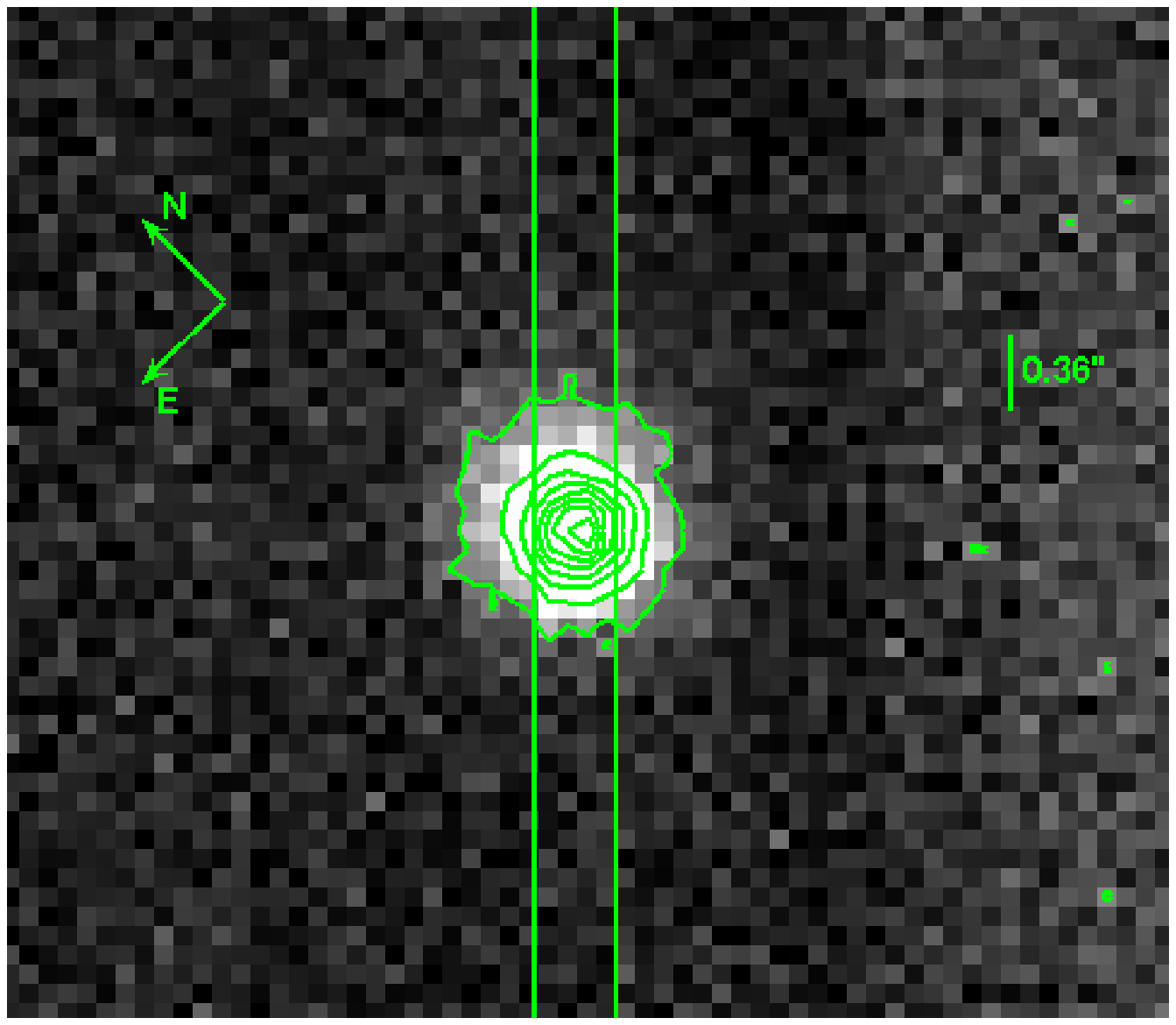}}&
   \subfigure[ ]{\includegraphics[clip=true,width=7.4cm]{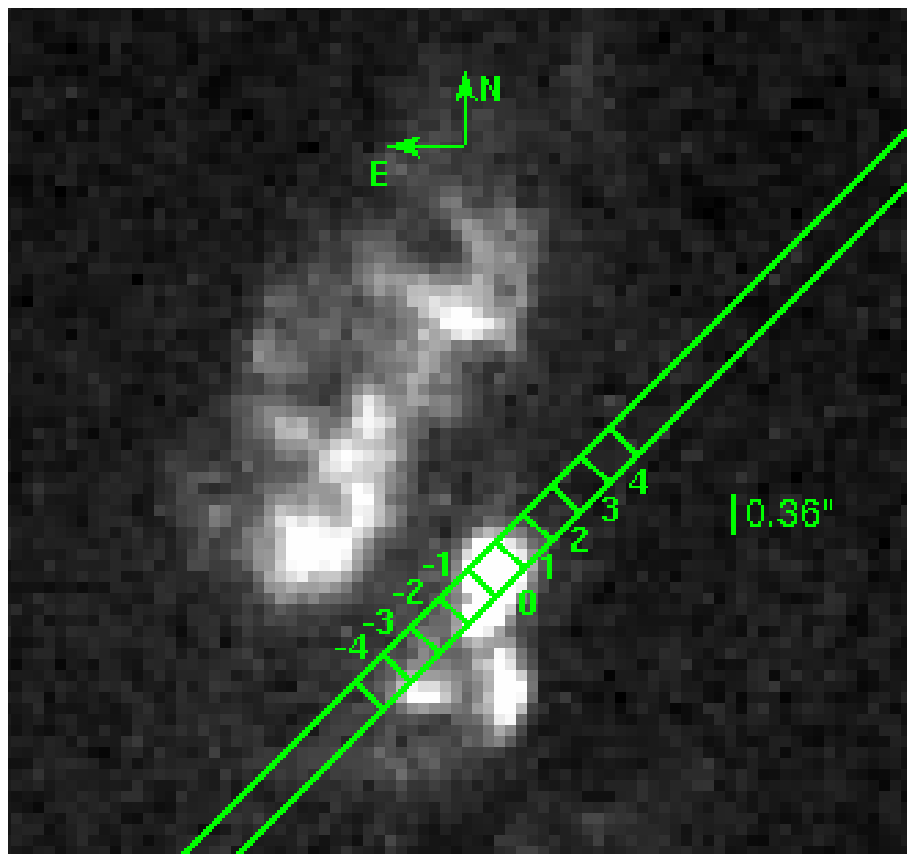}}\\
   \subfigure[ ]{\includegraphics[clip=true,width=8cm]{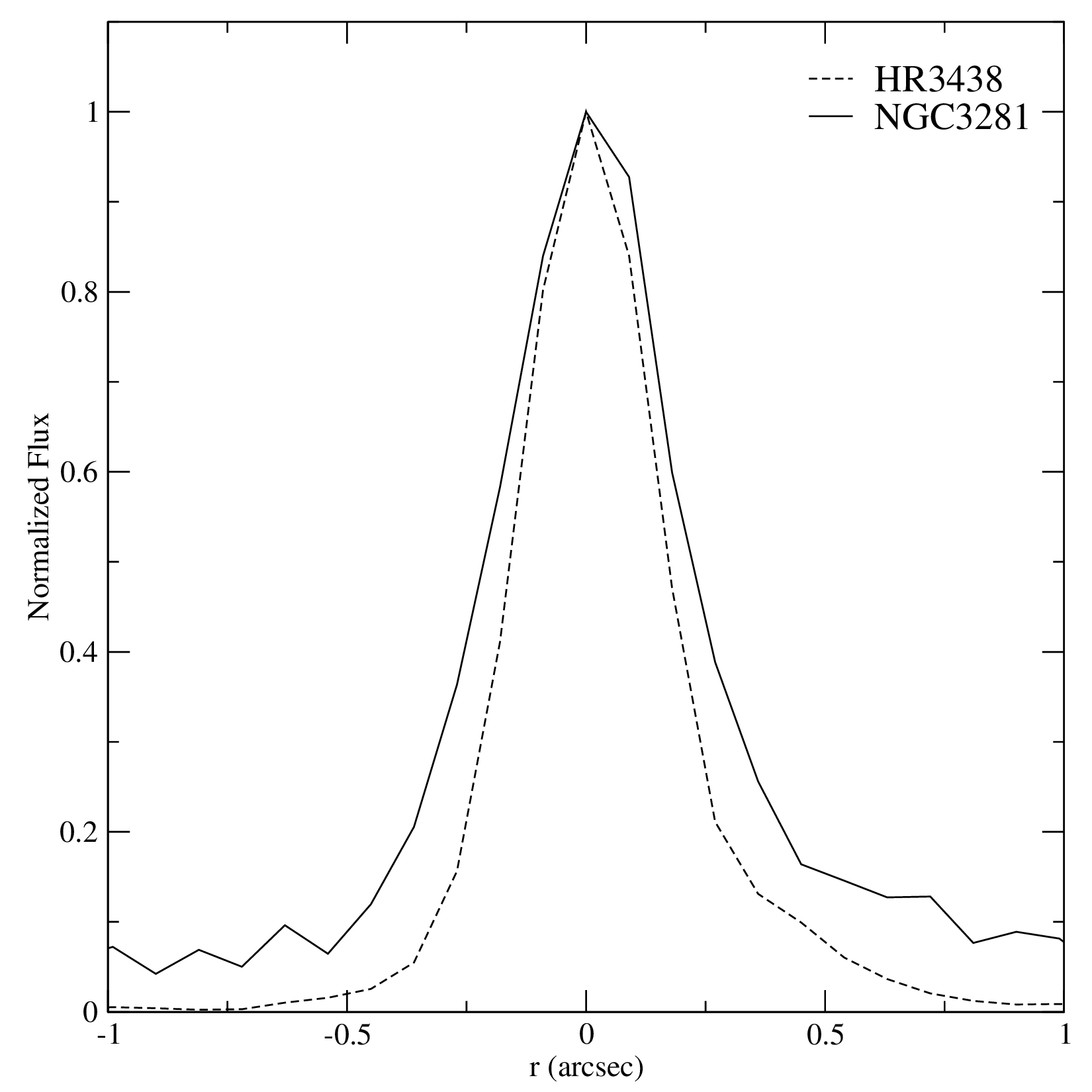}}\\
   \end{tabular}
   \par
  \end{centering}\vspace{-0.3cm}
\caption{(a) The T-ReCS long-slit position overplotted on the contoured NGC3281 acquisition image.
The N-band emission peak is centered in the image, which was supposed to coincide with the active nucleus. The contours are linear and stepped of 10\% of the peak.
(b) the slit superimposed on a [O\,{\sc iii}]$\lambda$5007\AA\ maps \citep{schmitt03b} and spectral extraction positions were labeled. 
(c) the T-ReCS N-band spatial emission profile, solid and dotted lines represent the NGC\,3281 galaxy and the HR3438 standard star, respectively. The fluxes are normalized to the peak value.}
\label{fenda}
\end{figure*}

Fig. \ref{fenda}a shows the slit superposed on the NGC\,3281 N-band acquisition image. 
Panel b of this figure illustrates the slit on the
narrow [O\,{\sc iii}] maps taken from \citet{schmitt03b}.
Panel c shows that the N-band galaxy luminosity profile
is extended with respect to that of the N-band stellar profile of HR3438.

\begin{figure*}
\centering
\includegraphics[scale=0.75,angle=0]{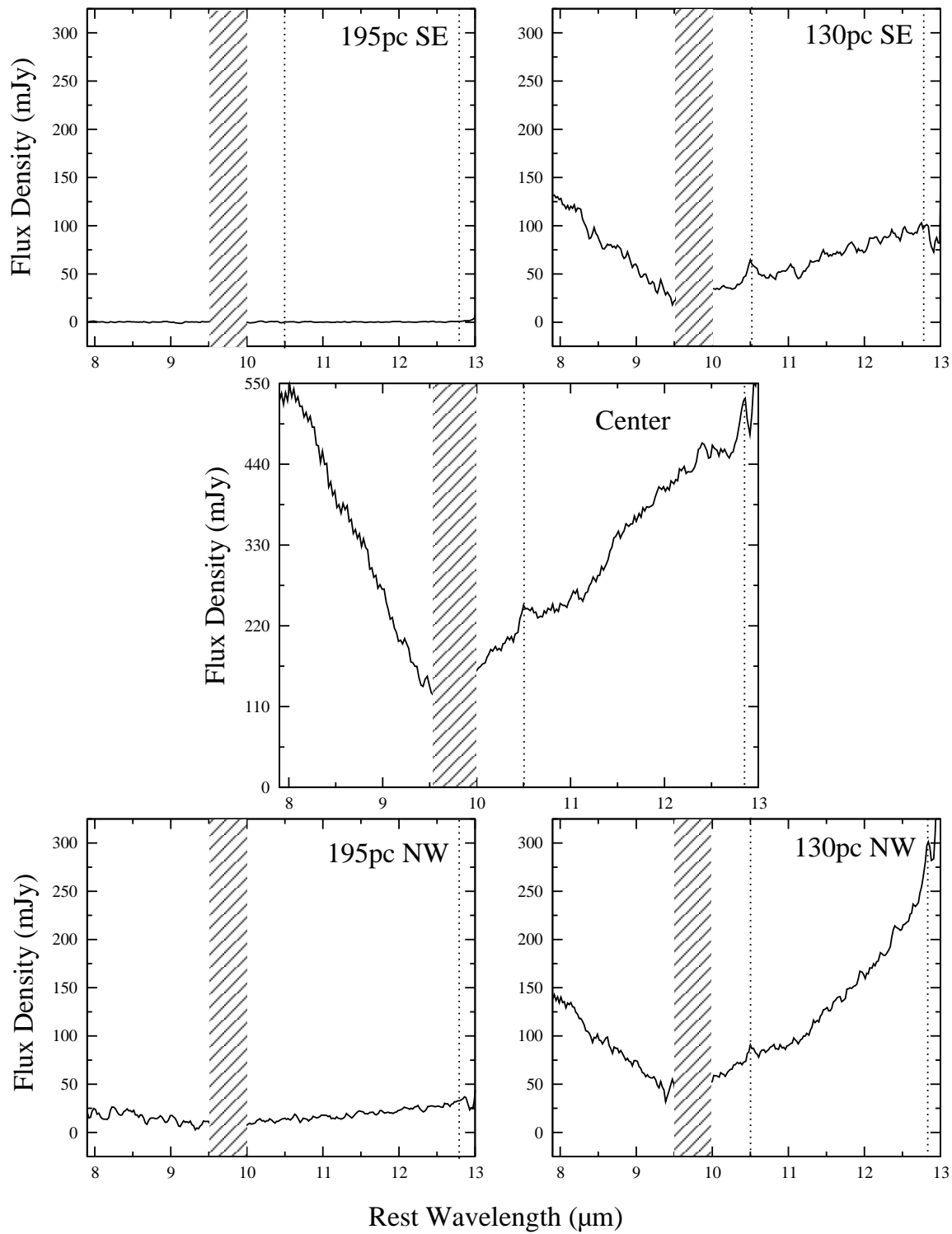}
\caption{Spectra of NGC\,3281 extracted in 65\,pc steps along P.A. = 315$^{\circ}$. 
Extraction positions are labeled, and dashed lines indicate the positions of the [S{\sc\,iv]}\,10.5\,$\mu$m and 
[Ne{\sc\,ii]}\,12.7\,$\mu$m ionic lines. Telluric O$_{3}$ band is represented by the hatched area.}
\label{spectra}
\end{figure*}

We extracted seven one dimensional (1D) spectra: one centered in the unresolved nucleus (4-pixel = 0.36 arcsec, which corresponds to 
$\sim$\,65 pc for a distance of 43 Mpc, using H$_{0}$ = 74 km s$^{-1}$ Mpc$^{-1}$), plus extractions centered at 130\,pc, 195\,pc and 260\,pc 
to the Northwest (NW) and Southeast (SE) directions. Deep silicate absorption is observed at the nucleus and at 130\,pc in both 
NW and SE directions of the nucleus (Fig. \ref{spectra}). The extraction at 195\,pc NW shows a weak absorption. The remaining extraction
does not present silicate absorption. Therefore, we can conclude that the dust is concentrated inside a radius of 200\,pc.

\section{Results and Discussion}\label{results}

\subsection{The Mid-IR Spectra}

The NGC\,3281 T-ReCS spectra (Fig. \ref{spectra}) clearly show that a deep silicate absorption, as well as the [S{\sc\,iv]}\,10.5\,$\mu$m and [Ne{\sc\,ii]}\,12.7\,$\mu$m emission lines, are 
observed up to 130 pc from the nucleus. In addition, PAH emission bands are not detected in the spectra, in agreement with the Spitzer spectrum \citep{wu09}. As discussed in the 
introduction, NGC\,3281 is one exception among Sy~2 galaxies, as most  ($\sim$\,80\%) of the objects in this class present PAH emission bands and no deep silicate absorption in their 
Mid-IR spectra \citep{sales10}.

In order to estimate the silicate absorption apparent strength ($S_{sil}$), we use the  definition of \citet{spoon07}:

\begin{equation}\label{sil}
S_{sil} = ln\,\frac{f_{obs}(9.7\,\mu m)}{f_{cont}(9.7\,\mu m)},
\end{equation}

where $f_{obs}$ is the observed flux density and $f_{cont}$ is the observed mid-line pseudo-continuum flux density. To determine the S$_{sil}$, we avoid telluric bands and emission line regions, using the same methodology as \citet{mason06}. The values derived from the four spatial positions where the silicate absorption is observed  are shown in column 7 of Tab.~\ref{lines}.

For the unresolved nucleus, we obtain S$_{sil}$ = $-1.3\,\pm\,0.1$. Similar values of S$_{sil}$ are observed in ULIRGs, while much weaker (average is -0.61) absorptions are seen in Sy\,2s \citep[see Fig.2 of][]{hao07}. The remaining three extraction  have similar values to that of the nucleus.

When dealing with dusty sources, a more physical parameter is the line of sight depth ($\tau$), which can be estimated from S$_{sil}$ at 9.7\,$\mu$m using the equation $ln(e^{-\tau_{9.7}}) = -\tau_{9.7} = \frac{f_{obs}(9.7\,\mu m)}{f_{cont}(9.7\,\mu m)}$ \citep{nenkova08b}. Furthermore, we can derive the apparent optical extinction using A$_{V}^{app}$ = $\tau_{9.7}\,\times$\,18.5\,$\pm$\,2 mag \citep{draine03}. The resulting values are shown in Tab.~\ref{lines}. The dust extinction  for the unresolved nucleus of NGC\,3281 is A$_{V}^{app}$ = 24\,$\pm$\,5\,mag, in agreement with the value (A$_{V}$ = 22\,$\pm$\,11 mag) derived by combined IR data \citep{simpson98}.
Peruse that our uncertainties is much lower than these authors.
We note from the values listed in Tab.~\ref{lines} that the position labeled as 1 (130\,pc from the nucleus at NW direction) presents a slightly higher extinction than the central one (A$_{V}$ = 28\,$\pm$\,8 mag).

To estimate in a more robust way the line of sight depth at 9.7\mc\ ($\tau_{9.7}$) in NGC\,3281 we used the {\sc pahfit}\footnote{Source code and documentation for {\sc pahfit} are available at http://tir.astro.utoledo.edu/jdsmith/research/pahfit.php} code \citep{smith07}, which assumes that the source spectrum is composed by continuum emission from dust and starlight, prominent emission lines, individual and blended PAH emission bands, and that the light is attenuated by extinction due to silicate grains. {\sc pahfit} models the extinction using the dust opacity law of \citet{kemper04}, where the infrared extinction is considered as a power law plus silicate features peaking at 9.7\,$\mu$m. For details see \citet{smith07}.

\begin{figure*}
\begin{centering}
\begin{tabular}{cc}
\subfigure[ ]{\includegraphics[scale=0.75,angle=0]{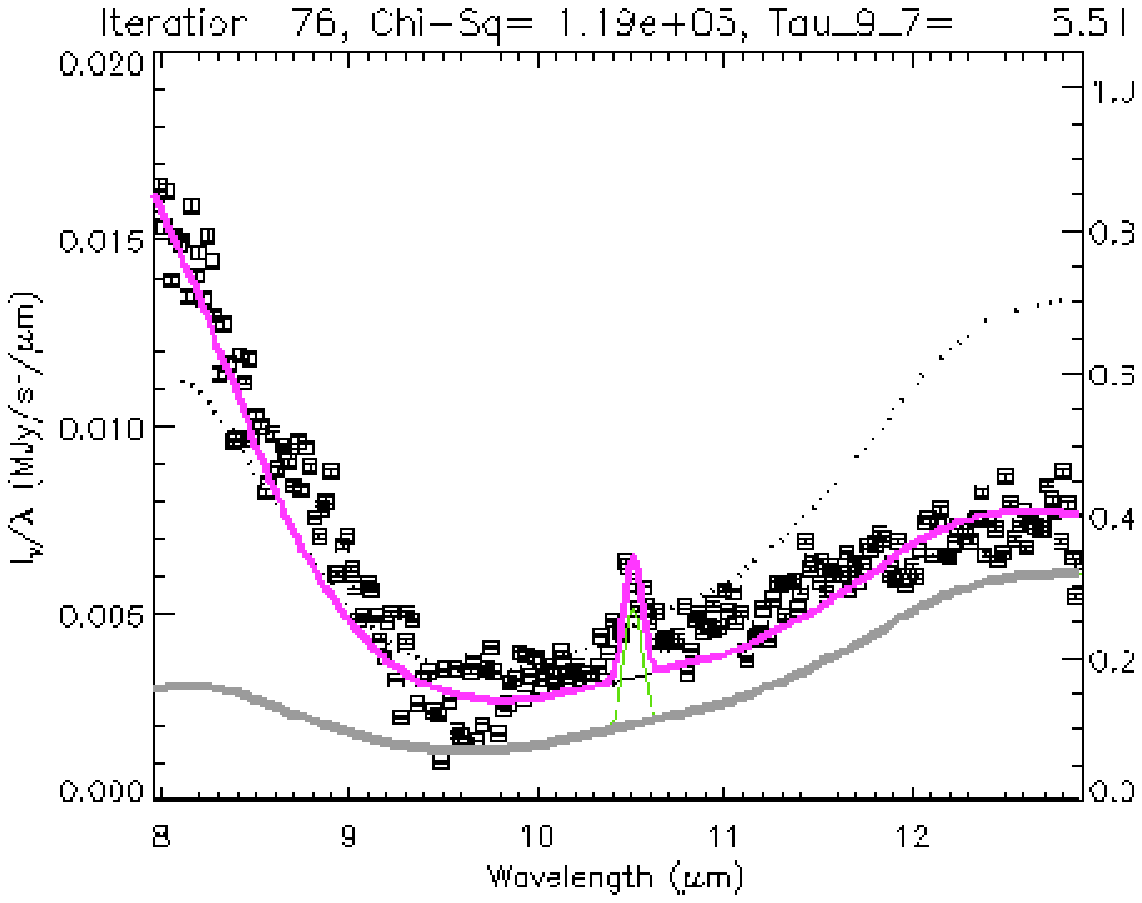}}&
\subfigure[ ]{\includegraphics[scale=0.75,angle=0]{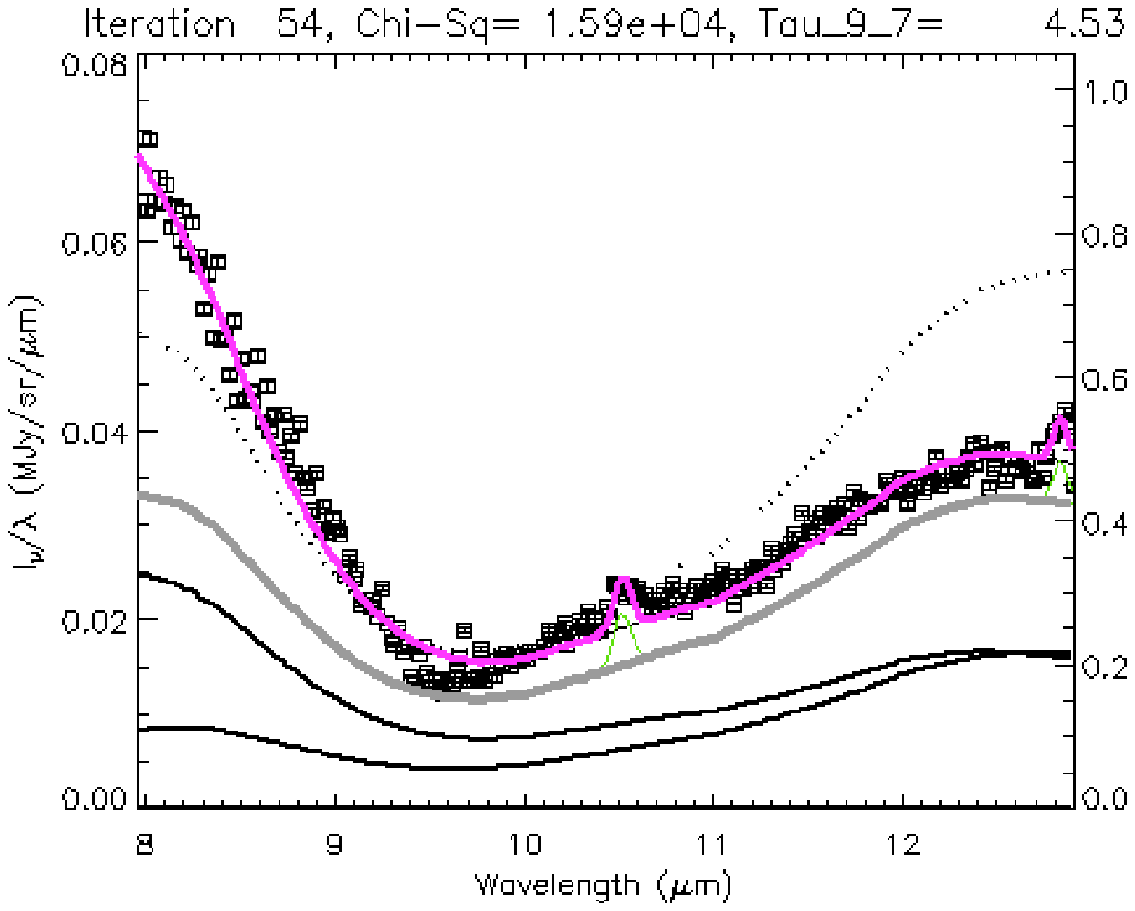}}\\
\subfigure[ ]{\includegraphics[scale=0.75,angle=0]{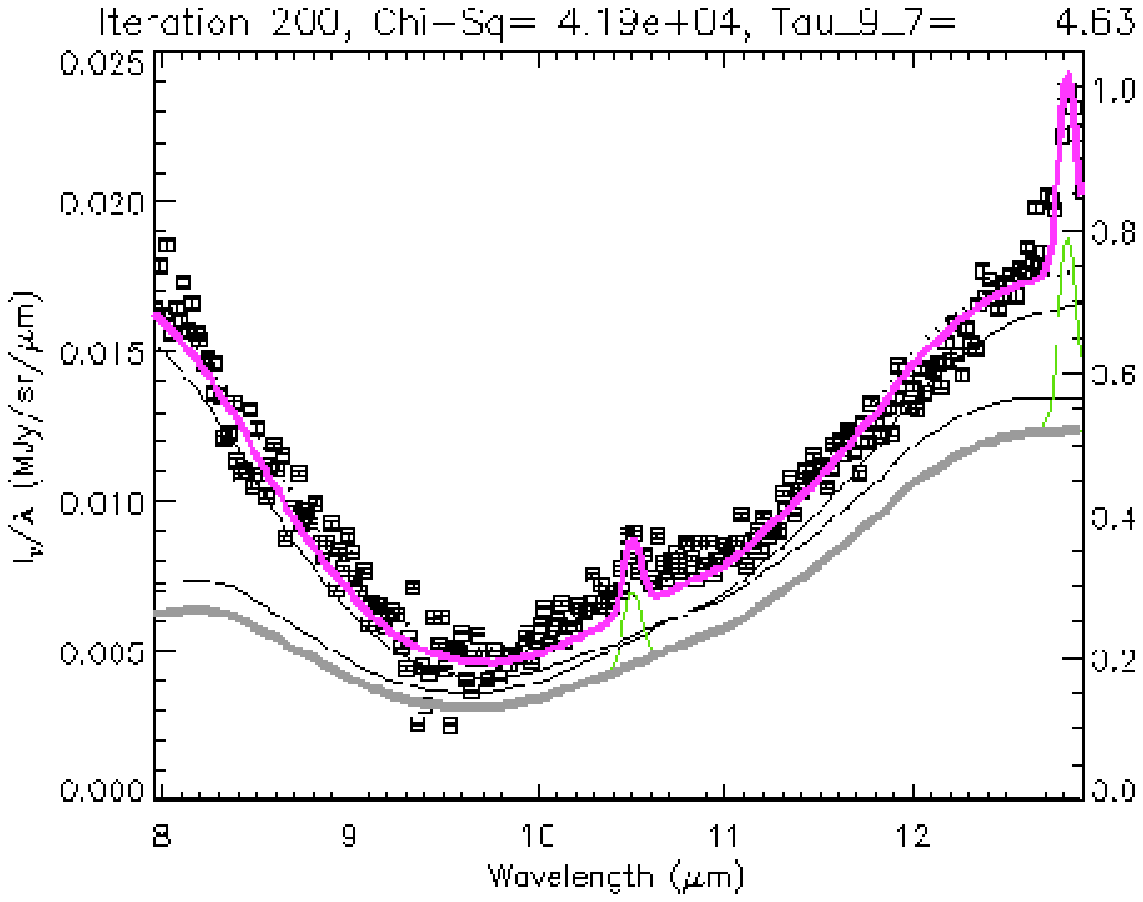}}&
\end{tabular}
\end{centering}
\caption{(a) The PAHFIT decomposition spectra of the 130\,pc SE (-1) extraction. (b) and (c) are the same for the unresolved nucleus (0) and 130\,pc NW (1) extractions, respectively.
Measurements are represented by squares, uncertainties are plotted as vertical error-bars, which are smaller than the symbol size.  Dotted black line indicates mixed extinction components.  
Solid grey and black line represent total and individual thermal dust continuum components, respectively. Green lines are emission lines. Magenta line represents the best fit model.}
\label{pahfit}
\end{figure*}

We used this approach to determine the contribution of the distinct
components to the spectral energy distribution (SED) of NGC 3281 as a
function of the distance to the galaxy centre. As the spectra do not
show PAH emission, we have not included this component in the fitting. The other input parameters are the same as those used by \citet{sales10}, which are appropriate for AGNs. The results of the decomposition of the NGC\,3281 spectra are shown in Fig.~\ref{pahfit}, and the derived silicate optical depths are given in Tab.~\ref{lines}. 

Clearly, $\tau_{9.7}$ varies along the slit, being $\tau_{9.7}$ = 4.5$\pm$0.7 at the unresolved nucleus and $\tau_{9.7}$ = 4.6$\pm$0.7 at 130\,pc NW extraction. In constrast, it 
reaches $\tau_{9.7}$ = 5.5$\pm$0.8 for the 130\,pc SE extraction.
The dust extinction (A$_{V}$) values estimated using the {\sc pahfit} derived line of sight depth, column 10 in Tab.~\ref{lines}, are much larger than those obtained from the S$_{sil}$. The latter are obtained from the
depth at 9.7\mc\, while the {\sc pahfit} code takes the whole profile of the silicate absorption feature into account, thus the S$_{sil}$ indicator should be taken as a lower limit for the A$_{V}$. The obtained opacity values were compared to those of the Seyfert sample from \citet{gallimore10}. The histogram of $\tau_{9.7}$ (Fig. \ref{histogram}) shows that the dust opacity in NGC\,3281 is one of the highest observed in Seyfert galaxies.

The thermal dust continuum components required to fit the observed spectra of the unresolved nucleus and 130\,pc SE 
is a black-body with T = 300\,K, while the 130\,pc NW spectrum requires it a black-body of T = 200\,K. This colder region has the deepest silicate absorption, consequently is more heavily obscured (A$_{V}$ = 102\,$\pm$\,26\,mag) than the nucleus and the SE region.
The presence of this obscuring material is consistent with the spatial distribution of the optical reddening found by \citet{thaisa92}. Note that our slit position is perpendicular to the cone observed in [O\,{\sc iii}]. See their Fig.~9 and 2.

In order to get additional support of that highly reddened region,
we overplot the T-ReCS slit position on the [O\,{\sc iii}]$\lambda$5007\AA\ maps taken from \citet{schmitt03b}. We found that the 130\,pc NW region coincide with a region where [O\,{\sc iii}] emission is not observed in agreement with those previously found from X-ray data. That is, the NGC\,3281 is a heavily obscured source.


\subsection{Is the observed silicate absorption detected at the unresolved nucleus associated with the dusty torus required by the unified AGN model?}

There are several torus models,
some authors take it as having an uniform dust density distribution \citep[e.g.][]{pier92,granato97,siebenmorgen04,fritz06}. However, for dust grains to survive in the torus environment, they should be formed by clumpy structures \citep{krolik88}, and new models have been developed where the dust grains are distributed in a clumpy medium \citep[e.g.][]{nenkova02,nenkova08a,nenkova08b,honig06}. Therefore, we compare the uncontaminated nuclear spectrum (see below) with theoretical SED obtained from Nenkova's models \citep[e.g.][]{nenkova02,nenkova08a,nenkova08b}.

In their modeling\footnote{The models are available from http://www.pa.uky.edu/clumpy/}, Nenkova et al. assume that the dusty material surrounding the AGN is distributed in clumps forming a toroidal geometry, and determine the following parameters: \textit{i})  the number of clouds \textit{$N_{0}$} in the torus equatorial radius; \textit{ii})  the optical depth for an individual cloud in the V band ($\tau_{V}$); \textit{iii}) the radial extension of the clumpy distribution, $Y=R_0/R_d$, where $R_0$ and $R_d$ are the outer and inner radius of the torus, respectively; \textit{iv})  the power law index $q$ that describes the radial density profile ($\propto r^{-q}$); \textit{v}) the torus angular width, constrained by a Gaussian angular distribution of $\sigma$; \textit{vi}) the observed viewing angle $i$.

\begin{figure*}
\centering
\includegraphics[scale=0.7]{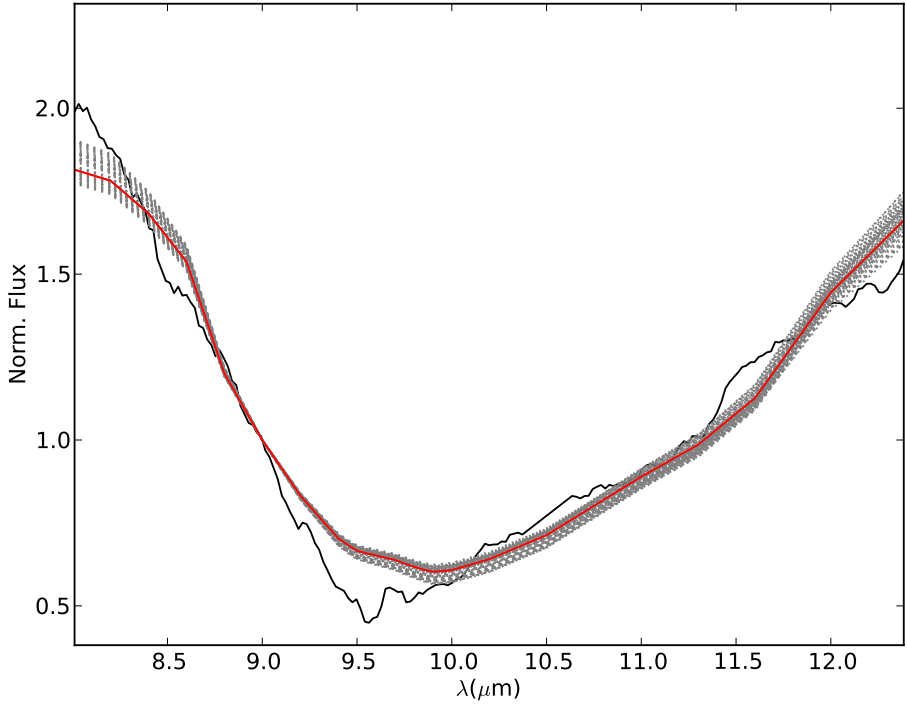}
\caption{Red and grey lines are the best and 10\% of best-fitting SEDs to the NGC\,3281 uncontaminated spectrum, respectively. All spectra were normalize in flux at 9.0$\mu$m.}
\label{nenkova}
\end{figure*}

\begin{figure*}
\centering
\includegraphics[scale=0.5,angle=-90]{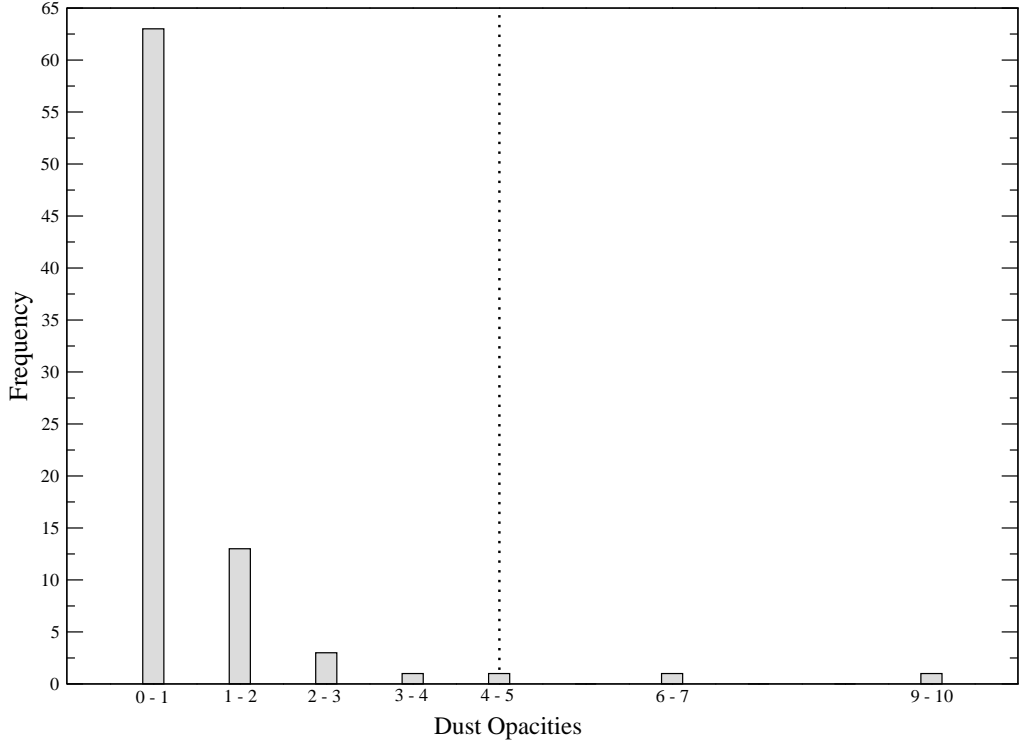}
\caption{Frequency histogram of the dust opacity of the 83 Seyfert galaxies taken from \citet{gallimore10}. Dotted line shows the value of the NGC\,3281.}
\label{histogram}
\end{figure*}

In order to properly analyze the spectrum from the nuclear dusty structure, it is necessary to first remove the host galaxy contribution. For this,  we have subtracted from the central aperture the spectrum obtained from averaging the two adjacent extractions (apertures -1 and 1, Fig.~\ref{fenda}). Noise effects are reduced by smoothing (with a rectangular box of size 5 pixels) the final spectrum. The emission lines were removed by simple interpolation and the region where the telluric band is located (Fig.\ref{spectra}) were not used in the fit. Then, the smoothed spectrum was compared with the $\rm \sim 10^6$ {\sc clumpy} theoretical SEDs, and the best fit is obtained by searching  for the minimum in the equation:

\begin{equation}
\chi^{2} = \frac{1}{N}\,\sum_{i=1}^{N}\,\left(\frac{F_{obs,\,\lambda_i} - F_{mod,\,\lambda_i}}{\delta_{\lambda_i}}\right)^{2},
\end{equation}
where N is the number of data points in the spectrum, F$_{obs,\,\lambda_i}$, F$_{mod,\,\lambda_i}$ and 
$\delta_{\lambda_i}$ are the 
observed, theoretical fluxes at each wavelength and their respective uncertainties. The later were taken as being 10\% of $F_{\lambda}$, which are the expected ones for T-ReCS observations \citep{radomski02,mason06}. Note that both F$_{obs,\,\lambda_i}$ and F$_{mod,\,\lambda_i}$ were normalized to unit at 9.0\mc, with the uncertainties being properly propagated. The parameter set that provides the best fit is shown in Table~\ref{statistic}, and the corresponding theoretical SED is over-plotted on the NGC\,3281 decontaminated nuclear spectrum (Fig.~\ref{nenkova}).
Due to the very large number of models required to cover the parameter space, we estimate the uncertainties for the best model following the approach of \citet[][see also \citet{nikutta09}]{mason09}, but instead of using only the three best models, we calculated the mean and standard deviation for the parameters considering all models with a $\chi^2$ value within 10\% of the best-fitting result (76 in total). The locus of the corresponding theoretical SEDs is plotted as a grey region in Figure~\ref{nenkova}. Our approach is similar to that used by \citet{nikutta09}, thus, we have also tested other $\chi^2$ deviation fractions (5\%, 10\%, 15\% and 20\%, corresponding to 17, 76, 151 and 291 acceptable solutions, respectively). In general our technique produce similar results than that obtained by \citet{nikutta09}, however, our methodology tends to produce a locus with less acceptable solutions than Nikutta's one. The mean parameters derived with the different $\chi^2$ deviations are shown in Table~\ref{statistic}.

The physical parameters of the best-fit model to the nuclear uncontaminated SEDs suggest that NGC\,3281 hosts a dusty toroidal structure. The dusty clouds, each with an optical depth 
$\tau_{V}$ = 40\,mag, occupy a toroidal volume with $R_0/R_d$ = 20. The distribution of the clouds follows a power-law radial behavior (r$^{-1.5}$), and the number of clouds along the equatorial
radius is 10. The angular distribution of the clouds is characterized by a width $\sigma$ = 70$^{\circ}$. These results are in agreement with the unified AGN model, which requires the
presence of a dusty structure that obscures the broad line region in Sy~2 galaxies. 
From the fitted models we would be looking in a direction  nearby of the  
torus equator ($i$ = 60$^{\circ}$)

In general, the torus parameters derived here are similar to those obtained by \citet{ramos09}, except for the number of clouds and optical depth per single cloud, where they find only 5 clouds with $\tau_{V}$ = 10 in the equatorial radius. This discrepancy can be due to the 
different constraints available, as \citet{ramos09} used Mid-IR photometric data \citep[see also][]{ramos11,alonso11}, while the data presented here would allow for a more detailed description of the 
spectral behavior of the silicate dust, thus providing a better input to the modeling procedure.

The \citet{nenkova08b} models adopt a standard dust-to-gas ratio, and assume the column density of a single cloud to be N$_{H}\,\sim\,10^{22} - 10^{23}\,cm^{-2}$. In addition, the total number of clouds at a viewing angle $i$ is given by a Gaussian distribution: N$_{obs}(i)$ = N$_{0}$\,exp$\left[-\left(\,\frac{90-i}{\sigma}\right)^2\right]$, thus resulting in a column density of N$^{(obs)}_{H}$ = \textit{N$_{obs}(i)$}\,N$_{H}$ at the observer view angle. Using the above, we can then derive a column density of N$_{H}\,\approx\,1.2\,\times\,10^{24}\,cm^{-2}$ for the uncontaminated nuclear spectrum of NGC\,3281, which is consistent with the values derived by \citet{vignali02} from X-ray observations ($\sim\,2\,\times\,10^{24}\,cm^{-2}$).

Moreover, \citet{levenson02} demonstrated that for heavily obscured AGN the iron K$\alpha$ EW values depend on the torus geometry. We compare the torus geometrical parameters derived in our paper with those of \citet{levenson02}, and found that the present torus aperture ($\sigma$ = 70$^{\circ}$) correspond to $\theta$ = 20$^{\circ}$ in the \citet{levenson02}'s models. The observer view angle has the same definition for both models (i = 60$^{\circ}$).
From Figure 2 of \citet{levenson02} and using the above parameters we estimate that the iron K$\alpha$ EW for NGC\,3281 is $\approx$ 2 - 3\,keV, which is in agreement with that found by \citet[][]{vignali02}, confirming their results, that this galaxy would shows a large EW for the iron K$\alpha$ line. We point out that this torus geometry was inferred from the silicate absorption line, this indicates that the X-ray absorbing column density, which makes NGC\,3281 a Compton-thick source, may also be responsible for the absorption at 9.7\,$\mu$m. Therefore, our results provide strong evidence that the silicate dust responsible for such absorption is located in the AGN torus. Following \citet{nenkova08b}, and adopting a dust sublimation temperature of 1500K, we estimate a torus outer radius of R$_{0}\,\sim$ 11\,pc. This value is consistent with the results of \citet{jaffe04,tristam07} and \citet{ramos09}.

We also derive the intrinsic bolometric luminosity for the AGN from the best-fit model, finding L$_{bol}$ = 1.9\,$ \times 10^{45}\,$erg s$^{-1}$.
This value agrees with the intrinsic luminosity obtained by \citet{vignali02} from X-ray data, when applying the conversion factor suggested by \citet{elvis94}, resulting L$_{X-ray}^{bol}$ $\approx\,3.2\,\times\,10^{44}\,$erg s$^{-1}$. We note that NGC\,3281 has a high X-ray luminosity when compared to objects studied by \citet{ramos09}.

\section{SUMMARY AND CONCLUSIONS}\label{conclusions}

In this work we present high-spatial resolution Mid-IR spectra of the
Compton-thick Sy~2 galaxy NGC\,3281. The ground-based Mid-IR data were
taken with the T-ReCS spectrograph, and our main results can be summarized as follows.

\begin{itemize}

\item NGC\,3281's spectra show a very deep silicate absorption, as well as
[S{\sc\,iv]}\,10.5\,$\mu$m and [Ne{\sc\,ii]}\,12.7\,$\mu$m ionic lines.
However, the spectra do not present PAH emission bands, which makes this galaxy
uncommon among the Sy~2s class \citep{sales10}. Furthermore, the dust is concentrated inside a radius of 200\,pc.

\item The optical extinction, A$_{V}$, of the nucleus and its neighborhood
was estimated using the S$_{sil}$ indicator and {\sc pahfit} code.
From the S$_{sil}$ indicator we infer the nuclear dust extinction
A$_{V}^{app}$ = 24\,$\pm$\,2\,mag, while the 
{\sc pahfit} code provides A$_{V}$ = 83\,$\pm$\,22\,mag.
We point out that the first value can only be used as a lower limit.
This large extinction confirms that this galaxy, if compared with 
\citet{gallimore10}' Seyfert sample (83 sources), is a heavily obscured 
source compatible with a Compton-thick galaxy scenario.

\item The black-body temperature due to dust continuum components
for the unresolved nucleus and the region at 130\,pc SE is T = 300\,K.
A colder black-body (T = 200\,K) was found for the
region at 130\,pc NW, which also has the highest derived visual
extinction (A$_{V}$ = 102\,$\pm$\,0.2\,mag).

\item The nuclear uncontaminated spectrum follows a 
clumpy torus model \citep{nenkova08b}, which suggests that NGC\,3281 has a dusty toroidal structure. 
The dusty clouds occupy a toroidal volume with $R_0/R_d$ = 20, with each cloud having an optical depth 
$\tau_{V}$ = 40\,mag. The distribution of the clouds follows 
a power-law radial behavior $r^{-1.5}$ and the number of clouds in the equatorial
radius is 14. The angular distribution of the clouds has width of $\sigma$ = 70$^{\circ}$. 
These results agree with the unified AGN model that to block the broad line region emission of Sy~2 galaxies,
requires a dusty structure. According to the best-fit model, we would be looking in the direction nearby of the 
torus equatorial radius ($i$ = 60$^{\circ}$), having an outer
radius of R$_{0}\,\sim$ 11\,pc.

\item We derive a column density of N$_{H}$ = 1.2\,$\times\,10^{24}\,cm^{-2}$ for NGC\,3281
indicates that the X-ray absorbing column density, which classifies NGC\,3281 as a Compton-thick source, may also be responsible for the absorption at 9.7\,$\mu$m. 
In addition, our results provide strong evidence that the silicate dust responsible for such absorption is located in the AGN torus. 
From Nenkova's models we find that the bolometric luminosity of this AGN 
is L$_{bol}$ = 1.9\,$ \times 10^{45}\,$erg s$^{-1}$, which shows that NGC\,3281 
is a very luminous source.

\end{itemize}

\begin{deluxetable}{lccccccccc}
\tabletypesize{\scriptsize}
\tablecaption{Line Measurements\label{lines}}
\tablewidth{0pt}
\tablehead{
\colhead{Position} & Label &\colhead{[S{\sc\,iv]}\,$_{10.5\,\mu m}$} & \colhead{EW[S{\sc\,iv]}} & \colhead{[Ne{\sc\,ii]}\,$_{12.7\,\mu m}$} & \colhead{EW[Ne{\sc\,ii]}} & \colhead{S$_{sil}$} & \colhead{A$_{V}^{app}$(mag)} & \colhead{$\tau_{9.7}$} & \colhead{A$_{V}$(mag)}
}
\startdata
130\,pc SE& -1 & 4.3\,$\pm$\,0.7 & 0.7& - & -& -1.4\,$\pm$\,0.1 & 26\,$\pm$\,5 & 5.5\,$\pm$\,0.8 & 102\,$\pm$\,26\\
Center &0& 6.2\,$\pm$\,1.0 & 0.1 & 1.3\,$\pm$\,0.3 & 0.02& -1.3\,$\pm$\,0.1 & 24\,$\pm$\,5 & 4.5\,$\pm$\,0.7& 83\,$\pm$\,22\\
130\,pc NW &1 & 2.3\,$\pm$\,0.4 & 0.2& 2.2\,$\pm$\,0.4 & 0.06& -1.5\,$\pm$\,0.3 & 28\,$\pm$\,8 & 4.6\,$\pm$\,0.7& 85\,$\pm$\,22\\
195\,pc NW& 2 & -& -& -& -& -1.1\,$\pm$\,0.1 & 21\,$\pm$\,4 & - &-\\
\enddata
\tablecomments{Columns 3 and 5 are values of the emission line integrated fluxes given in units of 10$^{-16}\,W\,$m$^{-2}$. 
Columns 4 and 6 are values of the equivalent widths in units of $\mu$m. Column 7 is the silicate strengths computed from eq. \ref{sil}.
Column 8 shows the apparent optical extinction values derived from S$_{sil}$ values. Column 9  
is the values of optical depth inferred by {\sc pahfit} and column 10 their optical extinction.
In order to determine the optical extinction values, we assume A$_{V}^{app}$/$\tau_{sil}$ = 18.5\,$\pm$\,2 \citep{draine03}.}
\end{deluxetable}

\begin{deluxetable}{lccccc}
\tabletypesize{\scriptsize}
\tablecaption{Fitted Parameters with CLUMPY Model\label{statistic}}
\tablewidth{0pt}
\tablehead{
\colhead{Parameter} & \colhead{Best Fit} &  \colhead{Average}& \colhead{Average}& \colhead{Average}&\colhead{Average}\\
\colhead{} & \colhead{} &  \colhead{5\%} & \colhead{10\%} & \colhead{15\%} & \colhead{20\%}\\
}
\startdata
Torus angular width ($\sigma$) & 	70$^{\circ}$ & 	68$^{\circ}$\,$\pm$\,3$^{\circ}$ & 62$^{\circ}$\,$\pm$\,6$^{\circ}$ &62$^{\circ}$\,$\pm$\,7$^{\circ}$ &61$^{\circ}$\,$\pm$\,8$^{\circ}$\\
Radial extent of the torus ($Y$) & 20& 20\,$\pm$\,0	& 23\,$\pm$\,4& 28\,$\pm$\,18&38\,$\pm$\,27\\
Number of clouds in the torus equatorial radius ($N_0$) & 14& 13\,$\pm$\,1	& 13\,$\pm$\,2&13\,$\pm$\,2&13\,$\pm$\,2\\
Power-law index of the radial density profile ($q$) & 	1.5&	1.5\,$\pm$\,0 &1.4\,$\pm$\,0.2&1.4\,$\pm$\,0.3&1.2\,$\pm$\,0.5\\
Observers' viewing angle ($i$) & 	60$^{\circ}$& 69$^{\circ}$\,$\pm$\,11$^{\circ}$ &	76$^{\circ}$\,$\pm$\,11$^{\circ}$&76$^{\circ}$\,$\pm$\,12$^{\circ}$& 72$^{\circ}$\,$\pm$\,18$^{\circ}$\\
Optical depth for individual cloud ($\tau_V$) & 	40\,mag & 40\,$\pm$\,0\,mag	& 40\,$\pm$\,0\,mag&40\,$\pm$\,3\,mag&41\,$\pm$\,8\,mag\\
Total number of solution & 	 & 17	& 76& 151& 291\\
\enddata
\tablecomments{Columns 2, 3, 4, 5 and 6 show the best value and the average for 5\%, 10\%, 15\% and 20\% of variation of the $\chi^2$, respectively.}
\end{deluxetable}

\acknowledgments

We thank an anonymous referee for helpful suggestions.
Based on observations obtained at the Gemini Observatory, which is operated by the 
Association of Universities for Research in Astronomy, Inc., under a cooperative agreement 
with the NSF on behalf of the Gemini partnership: the National Science Foundation (United 
States), the Science and Technology Facilities Council (United Kingdom), the 
National Research Council (Canada), CONICYT (Chile), the Australian Research Council (Australia), 
Minist\'{e}rio da Ci\^{e}ncia e Tecnologia (Brazil) 
and Ministerio de Ciencia, Tecnolog\'{i}a e Innovaci\'{o}n Productiva (Argentina).
We thank E. Balbinot, as well as Charles Bonatto, and A.C.C. acknowledges support from CNPq (grant 308985/2009-5) e Fapesp (grant 2010/19029-0).

\end{document}